\documentclass[epj]{webofc}
\usepackage[varg]{txfonts}
\usepackage{bm}
\begin{document}
\title{Hadronic structure of the photon at small $\bm{x}$ in holographic QCD}
\author{\firstname{Akira} \lastname{Watanabe}\inst{1,2}\fnsep\thanks{\email{watanabe@cycu.edu.tw}} \and
        \firstname{Hsiang-nan} \lastname{Li}\inst{2}\fnsep\thanks{\email{hnli@phys.sinica.edu.tw}}
}
\institute{Department of Physics and Center for High Energy Physics, Chung-Yuan Christian University, Chung-Li 32023, Taiwan, Republic of China
\and
Institute of Physics, Academia Sinica, Taipei 11529, Taiwan, Republic of China
}

\abstract{%
We present our analysis on the photon structure functions at small Bjorken variable $x$ in the framework of the holographic QCD.
In the kinematic region, a photon can fluctuate into vector mesons and behaves like a hadron rather than a pointlike particle.
Assuming the Pomeron exchange dominance, the dominant hadronic contribution to the structure functions is computed by convoluting the probe and target photon density distributions obtained from the wave functions of the U(1) vector field in the five-dimensional AdS space and the Brower-Polchinski-Strassler-Tan Pomeron exchange kernel.
Our calculations are in agreement with both the experimental data from OPAL collaboration at LEP and those calculated from the parton distribution functions of the photon proposed by Gl\"uck, Reya, and Schienbein.
The predictions presented here will be tested at future linear colliders, such as the planned International Linear Collider.
}
%
\maketitle
%
\section{Introduction}
\label{introduction}
A solid understanding of the nature of a photon has been one of the most longstanding problems in high energy physics for several decades.
A photon can fluctuate into quark-antiquark pairs or vector mesons in high energy scattering processes, and this feature provides us with an opportunity to investigate its partonic structure.
Experimentally, it can be studied via electron-photon deep inelastic scattering (DIS), and historically the data of photon structure functions measured at LEP have significantly deepened our knowledge of the energetic photon.
If the scattering is hard, perturbative QCD approaches are basically applicable because the pointlike component of the photon gives a dominant contribution, and a lot of studies have successfully been done so far, which strongly support the predictive power of them in high energy QCD processes.
However, in the small Bjorken $x$ region, the situation becomes quite different, because the photon behaves like a hadron rather than a pointlike particle.
When $x < 0.1$, we cannot neglect the hadronic contribution, and it becomes dominant at $x < 0.01$.
Since the photon consists of so many gluons with tiny momenta in this kinematic region, approaches motivated by effective models are practically the only way to investigate its nature.
Here we present our analysis on the photon structure functions at small $x$ in the framework of the holographic QCD, one of the effective models of QCD, based on our recent paper~\cite{Watanabe:2015mia}.

The AdS/CFT correspondence (more generally, gauge/string correspondence) has broadly gathered theoretical interests due to its strong predictive power for physical quantities in strongly coupled systems, while the nontrivial correspondence itself has not yet been mathematically proved.
In particular, its application to QCD processes is called holographic QCD, and it has been enthusiastically studied for over a decade.
As for DIS, after Polchinski and Strassler first performed the string calculations and discussed the Callan-Gross relation for the structure functions~\cite{Polchinski:2002jw}, a lot of elaborated studies have been done so far.
Brower, Polchinski, Strassler, and Tan (BPST) focused on the small $x$ region, and proposed the Pomeron exchange kernel~\cite{Brower:2006ea}, which gives a contribution of the Pomeron exchange to the total cross section.
Utilizing the BPST kernel, various phenomenological investigations have been done~\cite{Brower:2010wf,Watanabe:2012uc,Watanabe:2013spa,Watanabe:2015mia}, and their successes support further applications for high energy scattering processes in the kinematic region, where the Pomeron exchange gives a dominant contribution.

In this study, assuming the Pomeron dominance, we express the quasi-real photon structure functions by convoluting the BPST Pomeron exchange kernel and density distributions of the probe and target photons in the five-dimensional AdS space.
In contrast to the case of the nucleon structure functions, both the probe and target are pointlike particles, whose density distributions are expressed by the wave functions of the 5D U(1) vector field.
Also, two of three adjustable parameters of the model have been fixed in previous studies of the nucleon structure functions, and the remaining one is an overall factor which controls the magnitude of the structure functions.
Therefore we can test the predictive power of the model involving the BPST kernel through this study.

It will be demonstrated that our predictions for the photon structure functions are in agreement with the experimental data measured by OPAL collaboration at LEP~\cite{Abbiendi:2000cw}, although available data are limited in the small $x$ region.
We will show that our calculations are also consistent with those calculated from the parton distribution functions (PDFs) of the photon proposed by Gl\"uck, Reya, and Schienbein~\cite{Gluck:1999ub}, and discuss differences between the results of this work and previous studies of the nucleon structure functions.
The predictions particularly for the very low $x$ region will be tested at future linear colliders, e.g., the planned International Linear Collider.

\section{Model setup}
\label{model_setup}
\begin{figure}
\centering
\includegraphics[width=0.53\textwidth,clip]{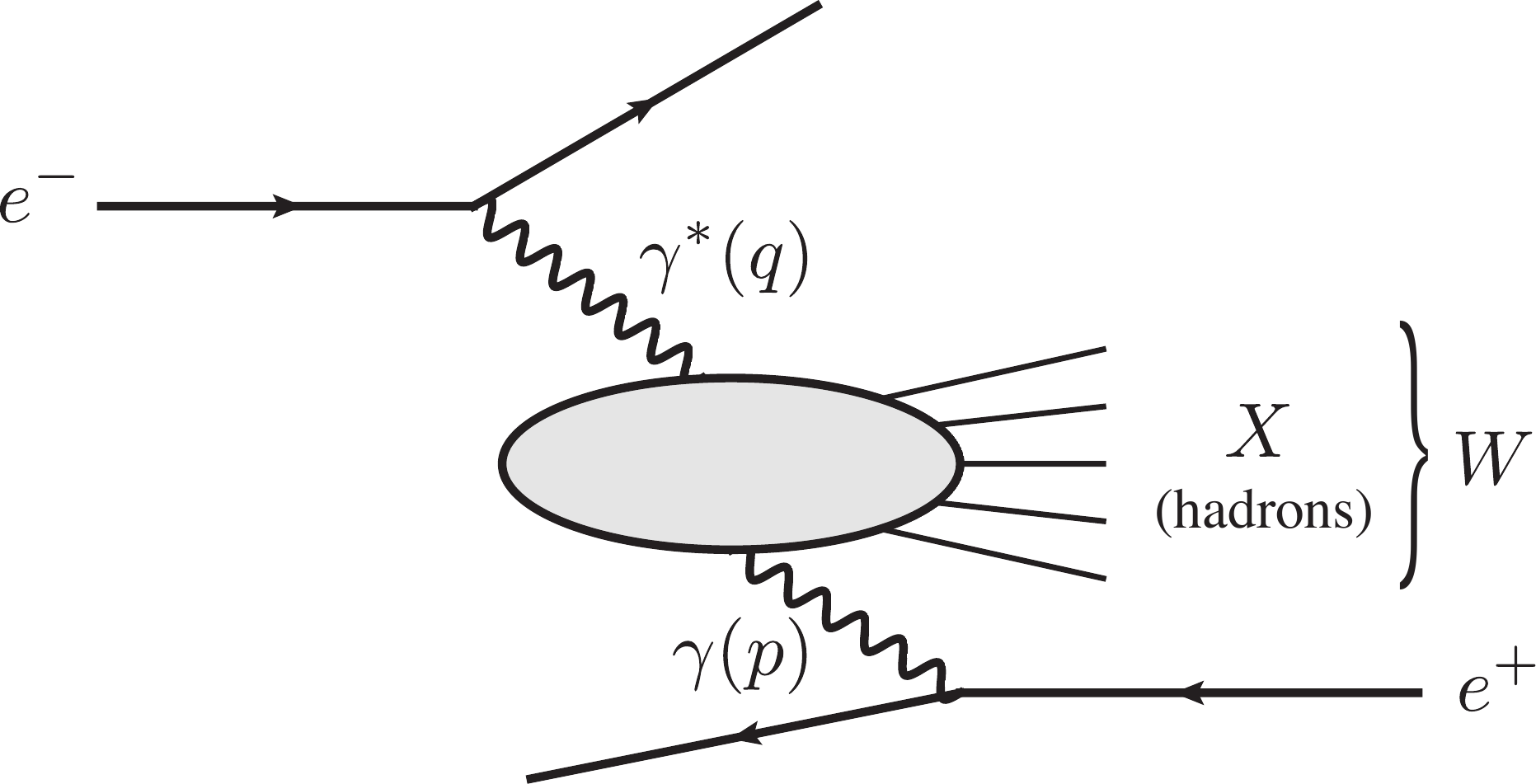}
\caption{Deeply inelastic electron-photon scattering process.}
\label{electron-photon_DIS}
\end{figure}

The electron-photon DIS is described in Fig.~\ref{electron-photon_DIS},
where $q$ and $p$ are the four-momenta of the probe and target photons, respectively, and $W$ denotes the invariant mass of the hadronic final state $X$.
The differential cross section of this scattering process is given by
\begin{equation}
\frac{{{d^2}{\sigma _{e\gamma  \to eX}}}}{{dxd{Q^2}}} = \frac{{2\pi {\alpha ^2}}}{{x{Q^4}}}\left[ {\left\{ {1 + {{\left( {1 - y} \right)}^2}} \right\}F_2^\gamma \left( {x,{Q^2}} \right) - {y^2}F_L^\gamma \left( {x,{Q^2}} \right)} \right], \label{differential_cross_section}
\end{equation}
where $\alpha$ is the fine structure constant, and $y$ is the inelasticity.
The Bjorken scaling variable $x$ is written by
\begin{equation}
x = \frac{Q^2}{Q^2 + W^2 + P^2},
\end{equation}
with $Q^2 = -q^2$ and $P^2 = -p^2$.
In this study, we focus on the kinematic region, where $W^2 \gg Q^2 \gg P^2$, so that the Bjorken variable can be approximated as $x \approx Q^2/W^2$.

With the BPST Pomeron exchange kernel, the two structure functions, $F_2^\gamma (x,Q^2)$ and $F_L^\gamma (x,Q^2)$, are expressed as~\cite{Watanabe:2015mia,Brower:2007qh,Brower:2007xg}
\begin{equation}
F^{\gamma}_{i} (x,Q^2) = \frac{\alpha g_0^2 \rho^{3/2}Q^2}{32 \pi ^{5/2}}
\int dzdz' P_{13}^{(i)} (z,Q^2) P_{24}(z',P^2) (zz') \mbox{Im} [\chi_{c} (W^2,z,z')], \label{Fic}
\end{equation}
where $i = 2,L$, and $g_0^2$ and $\rho$ are adjustable parameters which govern the magnitude and the energy dependence of the structure functions, respectively.
$P_{13}$ and $P_{24}$ are density distributions of the incident and target particles, respectively, in the five-dimensional AdS space.
To specify them, we consider the wave functions of the massless 5D U(1) vector field, which satisfies the Maxwell equation in the background spacetime and couple to leptons at the ultraviolet (UV) boundary~\cite{Polchinski:2002jw}.
We utilize it for the probe photon density distribution, then $P_{13}^{(i)} (z,Q^2)$ can be expressed as
\begin{align}
P_{13}^{(2)} (z,Q^2) &= Q^2 z \left[ K_0^2(Qz) + K_1^2(Qz) \right], \label{P13_2} \\
P_{13}^{(L)} (z,Q^2) &= Q^2 z K_0^2(Qz). \label{P13_L}
\end{align}
The first (second) term on the right-hand side of Eq.~\eqref{P13_2} represents the longitudinal (transverse) component of the wave function, and $K_0 (Qz)$ and $K_1 (Qz)$ are the modified Bessel functions of the second kind.
In this study, we concentrate on the unpolarized DIS process, so $P_{24} (z',P^2)$ takes the same functional form as $P_{13}^{(2)} (z,Q^2)$.
As to the four-momentum squared of the target photon, we focus on the real photon case, but $P^2$ cannot be exactly zero in practice.
Events with $P^2 \sim \mathcal{O} (0.01)$~GeV$^2$ have been measured~\cite{Nisius:1999cv,Krawczyk:2000mf}, so we assume a quasi-real photon with $P^2 = 0.01$~GeV$^2$ as the target particle here~\cite{Gluck:1999ub}.

The imaginary part of the conformal kernel was derived within the conformal field theory, and can be expressed as~\cite{Brower:2006ea,Brower:2007xg}
\begin{equation}
\mbox{Im} [\chi_{c} (W^2,z,z') ] = e^{(1-\rho)\tau} e^ {-[({\log ^2 z/z'})/{\rho \tau}]} / \sqrt{\tau}, \label{conformal_kernel}
\end{equation}
where $\tau$ is the conformal invariant defined by $\tau = \log (\rho z z' W^2 /2)$.
In the considered kinematic region, where $x \leq 10^{-2}$ and $Q^2 \leq 10$~GeV$^2$, nonperturbative hadronic contribution becomes important, so we also employ the modified kernel, whose imaginary part is given by~\cite{Brower:2006ea,Brower:2010wf}
\begin{align}
&\mbox{Im} [\chi_{mod} (W^2,z,z')] \equiv \mbox{Im} [\chi_c (W^2,z,z') ] + \mathcal{F} (z,z',\tau) \mbox{Im} [\chi_c(W^2,z,z_0^2/z') ],\label{modified_kernel} \\
&\mathcal{F} (z,z',\tau) = 1 - 2 \sqrt{\rho \pi \tau} e^{\eta^2} \mbox{erfc}( \eta ), \\
&\eta = \left( -\log \frac{zz'}{z_0^2} + \rho \tau \right) / {\sqrt{\rho \tau}}.
\end{align}
The second term on the right-hand side of Eq.~\eqref{modified_kernel} mimics the confinement effect in QCD, and suppresses the structure functions in the small $x$ region at fixed $Q^2$. The strength of this effect is controlled by the additional parameter $z_0$, whose inverse is of order of the QCD scale $\Lambda_{\rm QCD}$, and its smaller value gives a stronger suppression.
It should be also noted here that we treat the conformal and modified kernels as different models, and fix the parameter sets for them separately.

\section{Numerical results}
\label{numerical_results}
To evaluate the structure functions, we take the parameter sets obtained in Ref.~\cite{Watanabe:2013spa}.
In the previous studies on nucleon structure functions, since the target nucleon is an on-shell particle, the normalizable condition, $\int dz' P_{24}(z') = 1$, is satisfied.
However, the target quasi-real photon is a non-normalizable mode, so we have to newly fix the overall factor $g_0^2$ with experimental data.
We use 9 data points with $x \leq 0.025$ from the OPAL collaboration at LEP~\cite{Abbiendi:2000cw}, and obtain $g_0^2 = 17.51$ and 49.01 for the conformal and modified kernel cases, respectively.
When we perform the integration in Eq.~\eqref{Fic}, we need to avoid collecting unphysical contributions from the region, where the conformal invariant $\tau$ takes negative values.
To overcome this difficulty, we replace $z$ and $z'$ in the definition of $\tau$ with their average values defined by
\begin{equation}
{\bar{z}} \equiv \frac{{\int {dz} {z^2}{P_{13}^{(i)}} \left( {z,{Q^2}} \right)}}{{\int {dzz{P_{13}^{(i)}} \left( {z,{Q^2}} \right)} }}, \hspace{5mm} {\bar{z}'} \equiv \frac{{\int {dz'} {{z'}^2}{P_{24}} \left( {z',{P^2}} \right)}}{{\int {dz'z'{P_{24}}\left( {z',{P^2}} \right)} }},
\end{equation}
respectively.

\begin{figure}
\centering
\includegraphics[width=0.85\textwidth,clip]{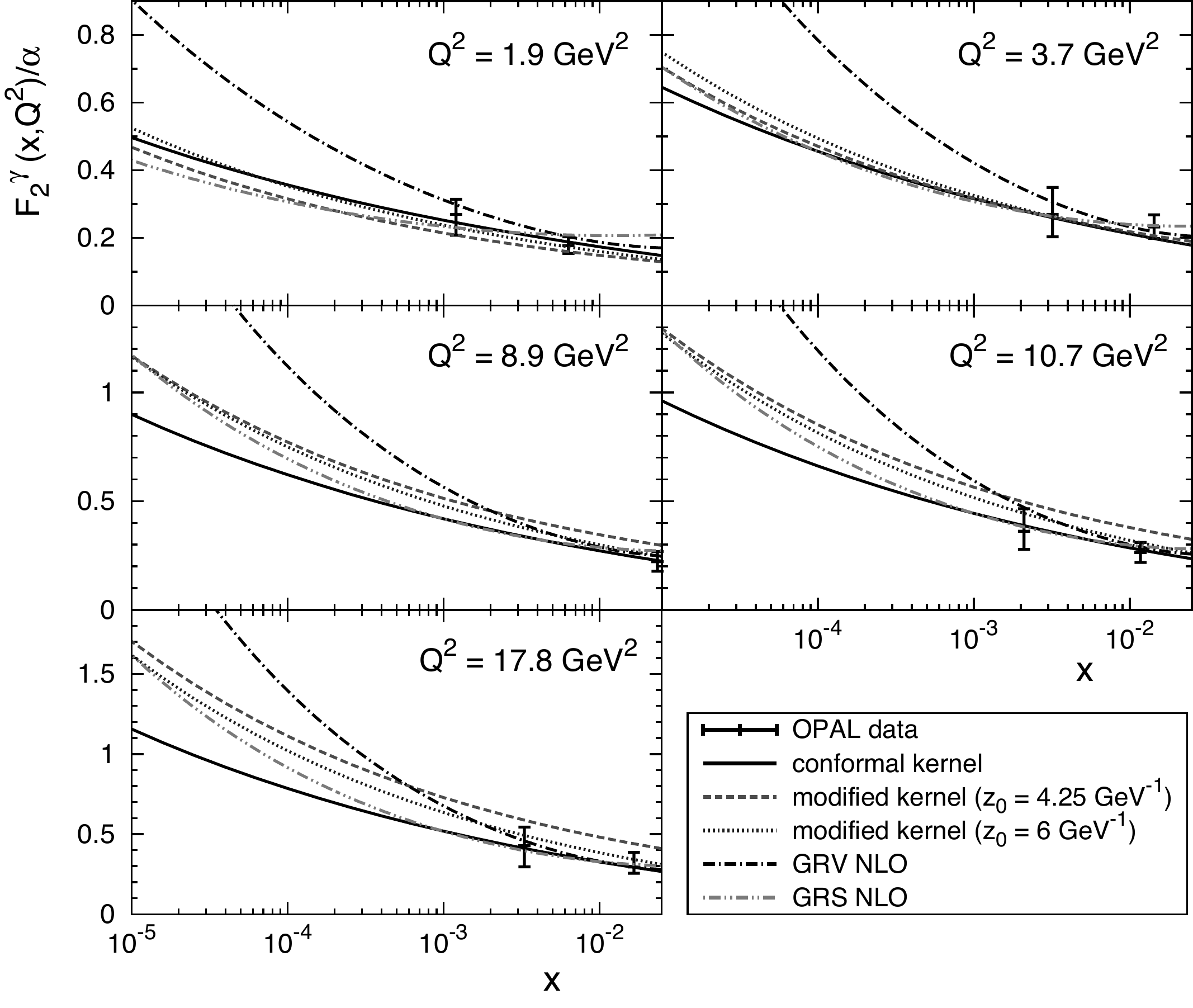}
\caption{$F_2^\gamma (x,Q^2)$ as a function of $x$ for various $Q^2$.
In each panel, experimental data from the OPAL collaboration at LEP~\cite{Abbiendi:2000cw} are denoted with error bars.
The solid, dashed, and dotted curves represent our calculations.
The dash-dotted and dashed double-dotted curves are from PDF sets, GRV~\cite{Gluck:1991jc}
and GRS~\cite{Gluck:1999ub}, respectively.}
\label{F2_photon}
\end{figure}

We display in Fig.~\ref{F2_photon}
our predictions of the photon structure function $ F_2$ with the conformal and modified Pomeron exchange kernels, experimental data from OPAL collaboration at LEP~\cite{Abbiendi:2000cw}, and those calculated from the well-known PDF sets, GRV~\cite{Gluck:1991jc} and GRS~\cite{Gluck:1999ub}, at next-to-leading-order accuracy.
Both GRV and GRS results include the charm-quark contribution in the kinematic region, where 
$W \geq 2 m_c $.
The hadronic contributions in the PDFs by GRV and GRS are parameterized by using the pion PDFs based on the vector meson dominance and the assumed similarity between the vector meson and the pion.
For the modified kernel results, we show two versions with different values of $z_0$, $z_0 = 4.25$~GeV$^{-1}$ and 6~GeV$^{-1}$, and the former value was obtained in Ref.~\cite{Watanabe:2013spa}.
One can see in Fig.~\ref{F2_photon} that all the curves are in agreement with the OPAL data.
In the smaller $x < 10^{-3}$ region, the $x$ dependence of the GRV results is obviously stronger than that of others.
The difference between GRV and GRS results was discussed in Ref.~\cite{Gluck:2001nx}.
Also in the region, $x$ dependence of the conformal kernel results is weakest for $Q^2 \geq 3.7$~GeV$^2$.
This can be understood via the results of $Q^2$ dependencies of the Pomeron intercept with the two kernels, which were shown in Refs.~\cite{Brower:2010wf,Watanabe:2012uc,Watanabe:2013spa}.
In those results, it is seen that the $Q^2$ dependence of the conformal kernel results is obviously weaker than that of the modified kernel results.
At fixed $x$, the suppression effect by the second term on the right-hand side of Eq.~\eqref{modified_kernel} becomes weaker with $Q^2$.

At $Q^2 \geq 8.9$~GeV$^2$, the conformal kernel results are more consistent with the experimental data than those of the modified kernel with $z_0 = 4.25$~GeV$^{-1}$, although it is known that the modified kernel can reproduce the nucleon structure function $F_2$ better compared to the conformal kernel~\cite{Brower:2010wf,Watanabe:2012uc,Watanabe:2013spa}.
We find that the modified kernel results can be improved by considering the relationship between the model parameter $z_0$ and the number of active quark flavors $f$.
In the nucleon case, its density distribution is localized in the infrared (IR) region, and there is almost no component near the UV boundary.
Due to this, heavy quark contributions are suppressed, and we get $f = 4$.
On the other hand, the density distribution of the quasi-real photon spreads broadly from the UV to IR region, which leads to the participation of all flavors, and we get $f = 6$.
Since the QCD scale $\Lambda_{\rm QCD}$ decreases with $f$, the parameter $z_0 \sim 1/\Lambda_{\rm QCD}$ for the quasi-real photon DIS should be larger than that for the nucleon DIS.
We choose $z_0 = 6$~GeV$^{-1}$ and $g_0^2 = 21.60$, and show the resulting $F_2^\gamma (x,Q^2)$ in Fig.~\ref{F2_photon}.
The results are perfectly consistent with the data within errors, implying that one of the fundamental QCD features is realized in the present model setup.

\begin{figure}
\centering
\includegraphics[width=0.95\textwidth,clip]{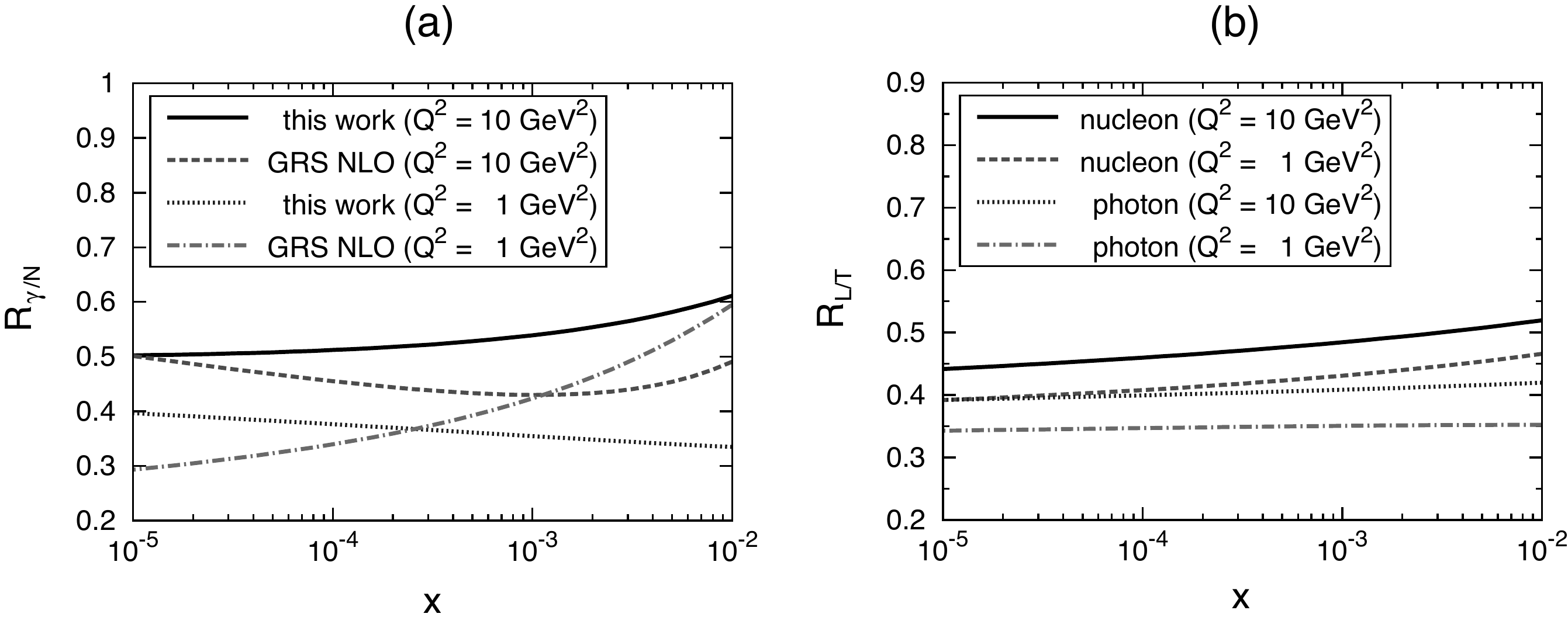}
\caption{Ratios (a) $R_{\gamma / N} = F_2^\gamma (x,Q^2) / [\alpha F_2^N (x,Q^2)]$ and (b) $R_{L/T} = F_L (x,Q^2) / F_T (x,Q^2)$ as functions of $x$ for $Q^2 = 1$ and 10~GeV$^2$.
The GRS results are calculated from the PDF set~\cite{Gluck:1999ub}.
$F_{i}^\gamma (x,Q^2)$ are obtained by employing the modified kernel with $z_0 = 4.25$~GeV$^{-1}$.
The nucleon results are taken from Ref.~\cite{Watanabe:2013spa}, and common to all the curves in the panel (a).}
\label{F2ratio_and_RLTvsx}
\end{figure}

Next we display in Fig.~\ref{F2ratio_and_RLTvsx}~(a)
the ratio $R_{\gamma / N} = F_2^\gamma (x,Q^2) / [\alpha F_2^N (x,Q^2)]$ to see the details of defferences between our and GRS results.
For $F_2^N (x,Q^2)$, we take the results in Ref.~\cite{Watanabe:2013spa}, which are in good agreement with the HERA data.
Both curves from our predictions are nearly linear and their slopes are small, which means the $x$ dependencies of the nucleon and photon structure functions $F_2$ are similar.
This may imply the universal feature of the BPST Pomeron exchange kernel.
Our and GRS results are close to each other at $Q^2 = 10$~GeV$^2$, because both results are in agreement with the data at this scale as shown in Fig.~\ref{F2_photon}.
However, at $Q^2 = 1$~GeV$^2$, the difference between two curves is much larger, and this difference can be seen in Fig.~\ref{F2_photon} also.
Particularly around $x \sim 10^{-2}$, the GRS results are nearly constant, while ours show the obvious $x$ dependence.

Finally, we investigate the longitudinal-to-transverse ratio of the structure functions defined as $R_{L/T} = F_L (x,Q^2) / F_T (x,Q^2)$, where $F_T = F_2 - F_L$, in Fig.~\ref{F2ratio_and_RLTvsx}~(b).
The photon results are compared with the nucleon results, which are taken from Ref.~\cite{Watanabe:2013spa}.
It is seen that the photon results are almost independent of $x$, but the $Q^2$ dependence is substantial.
Similar behaviors are observed for the nucleon results, which may imply universal features of the BPST Pomeron.

\section{Conclusion}
\label{conclusion}
In this study, we have investigated the electron-photon DIS, and calculated the structure functions $F_2$ and $F_L$ of the quasi-real photon in the kinematic region with the small Bjorken variable $x$ by the holographic QCD approach.
The photon structure functions are expressed by combining the BPST Pomeron exchange kernel and the density distributions of the probe and target photons in the five-dimensional AdS space, which can be described with the wave functions of the 5D U(1) vector field.
Our predictions are in agreement with the OPAL data, which supports the predictive power of this model.
We have also compared our results with those calculated from the well-known PDF sets, GRV and GRS, and found that ours are consistent with those from the latter one.
Since the hadronic component of the GRS PDF set is expressed with the pion PDFs, the consistency may imply the realization of the vector meson dominance in the present model setup.
Currently available experimental data at small $x$ are quite limited, but future linear colliders will enable us to further investigate the nature of a photon in the nonperturbative region.



\end{document}